\documentclass{article}
\usepackage{amssymb,amsmath,bm,graphicx,multirow}
\begin{document}
\begin{titlepage}
\noindent{\large\textbf{On the shape of a lightweight drop on a horizontal plane }}
\vspace{\baselineskip}
\begin{center}
Amir~H.~Fatollahi~{\footnote {fath@alzahra.ac.ir}}\\
\vspace{\baselineskip}
\textit{ Department of Physics, Alzahra University, Tehran
1993893973, Iran }
\end{center}
\vspace{\baselineskip}
\begin{abstract}
\noindent The shape of drop on a flat horizontal plane is obtained by including the first order of correction by the weight. The sphere solution of the weightless drop is used to introduce a new polar coordinate by which the perturbative expression for a region of a drop can be extended analytically to the entire surface of a drop having both the concave and the convex parts. Comparison with experimental data are presented.
\end{abstract}

\vspace{1cm}


\end{titlepage}

\section{Introduction}
To the interface of two mediums $a$ and $b$, it is assigned an energy per area of interface, the so-called interfacial energy coefficient $\gamma_{ab}$. For example, the liquid-vapor parameter $\gamma_\mathrm{lv}\equiv\gamma$ describes the energy content coming from the fact that liquid molecules near the surface have less neighbors than those in the bulk. The corresponding coefficient is called the surface tension. The shape of a drop of liquid on a solid surface, in the idealized case (absence of impurities and pinning effects), is determined by the quantities:
1) the surface tension $\gamma$, 2) the adhesion coefficient $\sigma$, 3) the shape of the solid surface, and due to the weight, 4) the drop's volume. The adhesion coefficient is defined by the surface tension, and the solid-liquid  and the solid-vapor  interfacial energies as \cite{gennes,bach}
\begin{equation}\label{1}
\sigma\equiv\gamma_\mathrm{sv}+\gamma-\gamma_\mathrm{sl}.
\end{equation}
At the solid-liquid-vapor point of contact, the contact angle $\vartheta$ in the equilibrium condition is given
by the Young equation
\begin{equation}\label{2}
\cos \vartheta =  \frac{\sigma}{\gamma} -1.
\end{equation}
\noindent The above relation determines $\vartheta$ for $0\leq \sigma \leq 2\gamma$. Three classes of possibilities for the contact angle are presented in Fig.~1.
The cases with $\sigma \approx 2\gamma $ and $ \sigma \ll \gamma$ correspond to the highest and the lowest spreading of the drop on the solid surface, respectively. Hence the name complete wetting befits to the case with $ \sigma \approx 2 \gamma$.

\begin{figure}[h]
\begin{center}
\includegraphics[width=0.8\columnwidth]{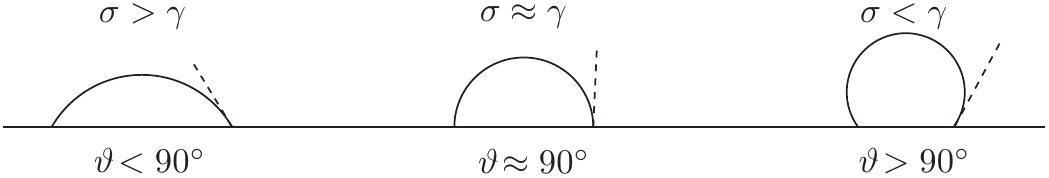}
\caption{Three classes of the drop's shape on a solid surface. }
\end{center}
\end{figure}

At every point on the drop surface the Young-Laplace relation holds \cite{gennes,bach}
\begin{equation}\label{3}
\Delta p =  \gamma \,\bigg (\frac{1}{R_1} + \frac{1}{R_2} \bigg)
\end{equation}
\noindent in which $\Delta p\equiv p_\mathrm{l}-p_\mathrm{v}$
is the difference pressure across the surface, and $(R_1,R_2)$ are two principal
radii of curvature of the surface at the point.
At each point of the drop's surface the total curvature $R_1^{-1}+R_2^{-1}$ is determined in terms of the surface equation and its derivatives. Provided by the hydrostatic laws, $\Delta p$ can be expressed in terms of the surface equation as well. So the Young-Laplace relation is the partial differential equation which, accompanied by appropriate boundary conditions, determines the shape of the drop's surface.

Equivalently, the shape of drop can be obtained by the minimizing the energy of a static system. While the surface tension tends to decrease the surface area of the drop, the adhesion coefficient tends to increase the surface area of the contact region, and the gravity tends to lower the center of mass of the drop. The competition between these effects determines the shape of the drop. For a drop with volume $V$ and density $\varrho$, one can give an estimation for each of the effects. The order of the drop's size is estimated by $L=V^{1/3}$. In many practical cases the surface tension and the adhesion coefficient, though with opposite effects, may be considered at the same order, meaning $\gamma$ and $\sigma$ are comparable. So the contribution of the interfacial energies, which is proportional to the area, is estimated by $\gamma\,L^2$. The contribution of the weight to the energy is given by $\varrho V g \, L=\varrho g \,L^4$, with $g$ as the gravitational acceleration constant. Comparing these contributions, one can differentiate three regions:
\begin{itemize}
\item $ L \ll \sqrt {\gamma / \varrho g} $: the effect of weight is small;
\item $ L \sim \sqrt {\gamma / \varrho g} $: the weight and the interfacial energies have comparable effects;
\item $ L \gg \sqrt {\gamma / \varrho g } $: the effect of weight is dominant.
\end{itemize}
In other words, the comparison between the length $\ell\equiv \sqrt {\gamma /\varrho g}$ and $L$,
or equivalently the value of the Bond dimensionless parameter $L^2\varrho g/\gamma$ would determine the regime. For a weightless drop only the contribution from the surface tension exists. Minimizing the area for a fixed volume, the shape of the drop's surface turns out to be part of a sphere.

The problem of a drop on a horizontal surface with the effect of surface tension being balanced with gravity has been studied for more than a century. The early numerical solutions go back to 1883 \cite{adams}, with updates by different authors \cite{staicop, padday, hartland}.
Different perturbative treatments of the problem have been developed over the years,
among them are those by \cite{chester,ehrlich,shanah,smith} for small drops (small Bond number).
As large drops (or vanishing surface tension drops) are theoretically an infinitely large and thin film of liquid subjected to the boundary conditions at the outer edge, the limit of large Bond number falls and has been studied in the context of singular perturbation problems \cite{rienstra}. Based on the similarity between the truncated oblate spheroid and drop's shape, an approximated profile is suggested in \cite{ryley} for the shape of the drop.
In \cite{lehman} a new numerical treatment of the problem is given based on a variational method to minimize the total energy of the drop, by which the use of the tables by \cite{adams} is more direct than the earlier treatments. As another effort in this direction, in \cite{obrien} the singular perturbation technique is used to obtain the asymptotic expressions describing the shape of small sessile and pendant drops.
The study of the profiles of resting drops in different situations are particularly important for practical purposes. In fact, one of the most common methods to measure the surface tension of liquids is based on the matching between calculated drop's profiles and the measured drop's shapes.
Over the years, the optimization of matching methods between the calculated profiles and the experimental data on drop's profiles has been the subject of several research pieces \cite{maze, neumann,kwok}.

It is the purpose of this note to develop a perturbation method to include the effect of the gravity on the shape of the drop's surface. Here in particular the shape of drop is obtained by including the first order of correction by the weight. The sphere solution of weightless drop is used to introduce a new polar coordinate by which the perturbative expression for a region of a drop can be extended analytically to the entire surface of a drop having both the concave and the convex parts.

\section{The shape of weightless drop}
It is known that the shape of a weightless drop is a part of a sphere. A proof for this is, however, given here, mainly to introduce the mathematical tools for the perturbative method. In particular,  it is the sphere solution by which the formerly mentioned polar coordinate is defined.

It can be shown that the total curvature of the surface $z=f (x, y) $ is given by \cite{oprea}
\begin{equation}\label{4}
\frac{1}{R_1} + \frac{1}{R_2} =- \frac{(1 + f_y ^ 2) f_ {xx} + (1 + f_x ^ 2) f_ {yy} - 2 f_x f_y f_ { xy}}{(1 + f_x ^ 2 + f_y ^ 2) ^ {3 / 2}},
\end{equation}
\noindent in which $ f_x = \partial_x f $, $ f_y = \partial_y f $,
$ f_ {xx} = \partial ^ 2_x f $, $ f_ {yy} = \partial ^ 2_ {y} f $ and $ f_ {xy} = f_ {yx} = \partial_x \partial_y f $. In the case with cylindrical symmetry, $ f $ depends on the combination
$ \rho = \sqrt {x ^ 2 + y ^ 2} $. In this case we find
\begin{equation}\label{5}
\frac{1}{R_1} + \frac{1}{R_2} = \frac{1}{\rho}~\frac{\mathrm{d}}{\mathrm{d}\rho}\bigg(\rho \,
\frac {|f '|} {\sqrt {1 + f '^ {\, 2}}} \bigg),
\end{equation}
\noindent where $ f '= \displaystyle {\frac{\mathrm{d} f}{\mathrm{d} \rho}} $. Defining
\begin{equation}\label{6}
\psi (\rho) = \frac{f '}{\sqrt {1 + f'^ 2 }}
\end{equation}
\noindent by which $ \displaystyle {\psi '(\rho) = \frac{f''}{( 1 + f'^ 2) ^ {3 / 2 }}}$, we get:
\begin{equation}\label{7}
\frac{1}{R_1} + \frac{1}{R_2} =
\frac{1}{\rho} \frac{\mathrm{d} }{\mathrm{d} \rho} \big (\rho |\psi| \big).
\end{equation}

\begin{figure}[t]
\begin{center}
\includegraphics[width=0.4\columnwidth]{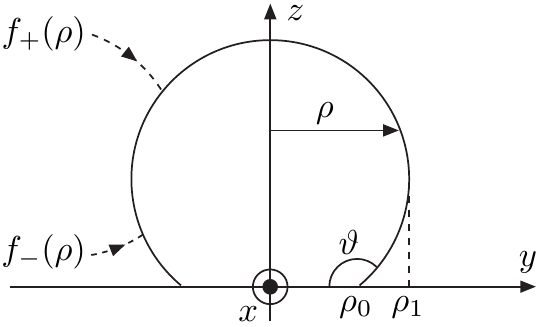}
\caption{The geometry of the mathematical setup for $\vartheta > 90^\circ$. }
\end{center}
\end{figure}

\noindent In absence of gravity ($g=0$), using (\ref{7}), we find for the Young-Laplace relation
\begin{equation} \label{8}
\frac{1}{\rho} \frac{\mathrm{d}}{\mathrm{d} \rho} \big (\rho |\psi_{0}| \big) = \frac{\Delta p_0}{\gamma }\equiv 2\, \kappa_0,
\end{equation}
\noindent in which $ \Delta p_0 $ is a constant, representing the difference pressure due to the surface tension of liquid in absence of gravity. In fact, due to the surface tension, the pressure inside the liquid drop is bigger than outside. So $\Delta p_0$, and thus $ \kappa_0 $, are positive.

In the case in which both concave and convex parts are present the entire surface can not be represented by only one function depending on $\rho$, simply because there are two different values of $z$ for some $\rho$ (Fig.~2). So the equation (\ref{8}) should be solved for the convex ($f_{0+}$) and concave ($f_{0-}$) parts separately  (Fig.~2). We mention $|f'_{0\pm}|=\mp f'_{0\pm}$. In the zero gravity case, however, since the right-hand side of equation (\ref{8}) is constant, one solution is related to the other simply by changing the sign of its derivative, or its corresponding $\psi_0$ by (\ref{6}). With the integration of (\ref{8})
\begin{equation}\label{9}
\mp\rho \psi_{0\pm} = \kappa_0 \rho ^ 2 + a_\pm
\end{equation}
\noindent where $ a_\pm $ are the constants of integration. By the above reasoning, in the present case $a_+=a_-$. Also $a_+$ should be set to zero in order that $\psi_{0+}$ does not blow up at $\rho=0$, for which by definition $|\psi(\rho)|\leq 1$. One then has
\begin{equation}\label{10}
\frac{f_{0\pm} '^2}{1 + f_{0\pm}'^ 2} = \kappa_0 ^ 2 \rho ^ 2
\end{equation}
\noindent and so
\begin{equation}\label{11}
\frac{\mathrm{d} f_{0\pm}}{\mathrm{d} \rho} = \mp \frac{\kappa_0 \rho}{\sqrt {1 - \kappa_0 ^ 2 \rho ^ 2}}
\end{equation}
\noindent for which by the integration we find:
\begin{equation}\label{12}
z = f_{0\pm} (\rho) = \pm \sqrt {\frac{1}{\kappa_0 ^ 2} - \rho ^ 2} + z_0
\end{equation}
\noindent which represents a sphere with radius $R=  \kappa_0 ^ {-1}>0 $ whose center is located on
$z$-axis at  $z=z_0$. As mentioned, in the case with $\vartheta\leq 90^\circ$ only the positive sign has meaning, and both $\pm$ signs should be kept in the case with $\vartheta>90^\circ$.

It is useful to check for the number of parameters involved. The parameters $\Delta p_0$, or equivalently $\kappa_0$, and $z_0$ are unknown at the first place. Following a simple geometrical argument in the sphere (see Fig.~2), we have
\begin{equation}\label{13}
\cos \vartheta = -\frac{z_0}{R},
\end{equation}
\noindent by which we have $z_0>0$ for $ \vartheta >90^\circ $  and $ z_0< 0 $ for $\vartheta< 90^\circ $, corresponding to center above and below the solid surface, respectively. So given by the relation for the volume
\begin{equation}\label{14}
V=\frac{\pi}{3} R^3\, (1-\cos\vartheta)^2(2+\cos\vartheta)
\end{equation}
\noindent $z_0$ and $R$ are fixed. The parameters $\rho_0$ and $\rho_1$ in Fig.~2, as the contact and equatorial radii, respectively, can be obtained once the equation of sphere being solved for $z=0$ and $z=z_0$, yielding
\begin{equation}\label{15}
\rho_0=R\sin\vartheta,~~~~~~~~~~ \rho_1=R.
\end{equation}
The place of drop's apex in the spherical solution is
\begin{equation}\label{16}
h_0=R+z_0
\end{equation}

Before to proceed, let us to introduce an identity for later uses.
The difference pressure in presence of gravity gets contribution
from the weight of the drop's layers as well. So we have for the ratio
\begin{equation}\label{17}
\frac{\Delta p (z)}{\gamma} =2 \kappa +  \frac{\varrho g}{\gamma} (h - f (\rho))
\end{equation}
in which $h$ is the height of the drop's apex, and $\kappa$, similar to its
counterpart $\kappa_0$, represents the difference pressure due to the
surface tension but here in presence of gravity. The contribution of the surface tension
to pressure in presence of gravity is different and simply can be
understood as the area of the drop is changed due to the effect of weight. Hence the surface
tension contribution to the energy of the drop is different due to the gravity. So, the
Young-Laplace  relation reads
\begin{equation}\label{18}
\mp\frac{1}{\rho} \frac{\mathrm{d}}{\mathrm{d} \rho} \big (\rho \psi_\pm \big) =
 2 \kappa +  \frac{\varrho g}{\gamma} (h- f_\pm).
\end{equation}
Integrating the above for the upper and lower parts gives the followings,
\begin{align}\label {19}
 \rho_1 &=    \left( \kappa +\frac{\varrho g}{2\gamma}h\right) \rho_1 ^ 2 -
\frac {\varrho g} {\gamma} \int_ {0} ^ {\rho_1} \rho f_+(\rho) \mathrm{d} \rho \\
\label{20}
\rho_1- \rho_0 \sin \vartheta  & =  \left( \kappa +\frac{\varrho g}{2\gamma}h\right)
(\rho_1 ^ 2 - \rho_0 ^ 2) - \frac {\varrho g} {\gamma} \int_ {\rho_0} ^ {\rho_1} \rho f_-(\rho)\mathrm{d} \rho
\end{align}
in which we have used
\begin{equation}\label{21}
\frac  {f _-'} {\sqrt {1 + f_-'^{2}}} \bigg | _ { \rho = \rho_0} =
\frac {-\tan \vartheta } {\sqrt {1 + \tan ^ 2 \vartheta }} = \sin \vartheta
\end{equation}
for $\vartheta > 90^\circ$. Subtracting (\ref{19}) and (\ref{20}) gives
\begin{equation}\label{22}
\kappa +\frac{\varrho g}{2\gamma}h
= \frac{\sin \vartheta }{\rho_0} + \frac {\varrho g V} {2 \pi \gamma \,\rho_0^2}
\end{equation}
in which we have used the relation for the volume of drop,
\begin{equation} \label {23}
\frac  {V} {2 \pi}  = \int_0 ^ {\rho_1} \rho f_ +(\rho) \mathrm{d} \rho - \int_ {\rho_0} ^ {\rho_1} \rho f_- (\rho)\mathrm{d} \rho
\end{equation}
\noindent  It is easy to show that identity (\ref{22}) is valid for the acute contact angle
($\vartheta < 90^\circ$), as well. It is reminded that in obtaining (\ref{22}) no approximation is used, and so it is an exact relation.

\section{The shape of lightweight drop}
Here we consider the first correction of gravity to the shape of a drop, supposedly applicable to the drops with tiny weight, or equivalently small volume.
It is clear from (\ref{18}) that the effect of weight, as mentioned earlier, appears in the combination
$ \varrho g/ \gamma$. By the help of volume $V$, it is useful to introduce the Bond dimensionless parameter
$ \lambda\equiv V ^ {2 / 3}\varrho g/\gamma $.
This supposedly small parameter helps to develop a perturbative expansion for the contribution of gravity on the shape of drop. At the first order of correction one has
\begin{equation}\label{24}
z = f (\rho) = f_0 (\rho) + \lambda f_1 (\rho)
\end{equation}
\noindent where $ f_0 (\rho) $ is the sphere solution found in the previous section.
By inserting above in (\ref{6}) one finds for $\psi_+(\rho)$
\begin{equation}\label{25}
\psi_+(\rho)=\frac{f'_+}{\sqrt{1+f'^2_+}}=\frac{f_{0+}'}{\sqrt{1+f_{0+}'^2}}+\lambda\frac{f_{1+}'}{(1+f_{0+}'^2)^{3/2}}+O(\lambda^2).
\end{equation}
\noindent The combination $\kappa +\frac{\varrho g}{2\gamma}$ in the right-hand side of (\ref{18}) can easily be rearranged based on the effect of gravity using the identity (\ref{22}). As it is expected that the contact radius $\rho_0$ is changed under the effect of gravity, at the first order of perturbation in $g$, by replacing $\rho_0=R\sin\vartheta + \delta \rho_0$, the identity comes to the form
\begin{equation}\label{26}
\kappa +\frac{\varrho g}{2\gamma} h
\simeq \frac{1}{R}- \frac{\delta\rho_0}{R^2 \sin \vartheta }
+ \frac {\varrho g V} {2 \pi \gamma \,R^2 \sin^2\vartheta }
\end{equation}
We later will find that $\delta\rho_0>0$, as expected. By insertion (\ref{25}) and (\ref{26}) in (\ref{18}), and using the fact that $f_0$ satisfies the equation with $\lambda=0$ (eq.~(\ref{8})), for the case with $\vartheta<90^\circ$ or the upper half of case with $\vartheta>90^\circ$ the Young-Laplace  relation reads
\begin{equation}\label{27}
\frac{\mathrm{d}}{\mathrm{d} \rho}
\Big [\rho \frac{f_{1+} '}{(1 + f_{0+}'^2) ^ {3 / 2}} \Big] = \frac{1}{V^{2 / 3}}\rho\, (a+f_{0+} (\rho))
\end{equation}
\noindent in which
\begin{equation}\label{28}
a= \frac{2\gamma}{\varrho g} \frac{\delta\rho_0}{R^2 \sin \vartheta }
- \frac {V} { \pi R^2 \sin^2\vartheta }
\end{equation}
Using $ f_{0+} (\rho) = \sqrt {R ^ 2 - \rho ^ 2} + z_0 $, integrating (\ref{27}) from $0$ to $\rho$ gives
\begin{equation}\label{29}
\rho \frac{f_{1+} '}{(1 + f_{0+}'^ 2) ^ {3 / 2}} =\frac{1}{V ^ {2 / 3}}
\Big (\frac{1}{2} (z_0+a) \rho ^ 2 -
\frac{1}{3 } (R ^ 2 - \rho ^ 2) ^ {3 / 2} +\frac{1}{3}R^3 \Big).
\end{equation}
\noindent  Again using the expression for $ f_{0+} (\rho)$, we find
\begin{equation}\label{30}
f_ {1+} '(\rho) =\frac{R ^ 3}{V ^ {2 / 3}} \Big [
\frac{(z_0+a)\rho}{2(R ^ 2 - \rho ^ 2) ^ {3 / 2}} - \frac{1}{3 \rho}
+\frac{R ^ 3 }{3 \rho (R ^ 2 - \rho ^ 2) ^ {3 / 2}} \Big].
\end{equation}
\noindent It is useful to note that for the above expression, the limit $\rho\to 0$ exists and is zero, as expected. However, we mention that the above expression diverges in the limit $\rho\to R$, indicating that the present form of the perturbative solution fails for drops with contact angle close to or greater than $90^\circ$. We will come back to this issue later.
The above expression can be used to find the corrected value of the contact radius $\rho_0$, or equivalently $\delta\rho_0$. Subjected to the condition that the drop intercepts the solid surface with contact angle $\vartheta$, and also the constraint on the volume of drop, the expression (\ref{30}) should satisfy the followings:
\begin{align}\label{31}
-\tan\vartheta = f_+'(\rho_0) &\simeq f_{0+}'(\rho_0)
+\lambda f_{1+}'(R\sin\vartheta),  \\
\label{32}
V=\pi \int_0^{\rho_0} \rho^2 |f_+'(\rho)| \mathrm{d}\rho &\simeq
-\pi \int_0^{\rho_0}\!\! \rho^2 f_{0+}'(\rho) \mathrm{d}\rho
-\pi\lambda \int_0^{R\sin\vartheta}\!\! \rho^2 f_{1+}'(\rho) \mathrm{d}\rho
\end{align}
in which $\rho_0=R\sin\vartheta+\delta\rho_0$. We mention, due to presence of $\lambda$, it is sufficient to insert the unperturbed values of previous section in the expression for $ f_{1+}'$.
It is easy to check that, thanks to the identity (\ref{22}), the first in above is automatically satisfied. By the second condition the change in the contact radius is found to be
\begin{equation}\label{33}
\delta\rho_0= \frac{\varrho g R^3}{6 \gamma} \frac{(1-\cos\vartheta)^2}{\sin\vartheta (2+\cos\vartheta)}>0
\end{equation}
by which the constant $a$ is obtained
\begin{equation}\label{34}
a=-R\frac{(1-\cos\vartheta)(3+\cos\vartheta)}{3(2+\cos\vartheta)}.
\end{equation}
Once again the relation (\ref{30}) can be integrated, leading to
\begin{equation}\label{35}
f_ {1+} (\rho) =  \frac{R ^ 3}{V ^ {2 / 3}} \Big [
 \frac{3(z_0+a)+2R}{6 \sqrt { R ^ 2 - \rho ^ 2}} -
\frac{1}{3} \ln \Big (\frac{R + \sqrt {R ^ 2 - \rho ^ 2}}{2R} \Big)
\Big] + b
\end{equation}
\noindent where $ b $ is a constant, that should be determined by the condition
\begin{equation}\label{36}
0=f_{+}(\rho_0)\simeq  f_{0+}(\rho_0)+\lambda f_{1+}(R\sin\vartheta).
\end{equation}
Using above we find
\begin{equation}\label{37}
b= \frac{R ^ 3}{V ^ {2 / 3}} \Big [  \frac{\cos\vartheta}{2(2+\cos\vartheta)}+
\frac{2}{3} \ln \cos \frac{\vartheta}{2} \Big]
\end{equation}
In the case $\vartheta>90^\circ$, (\ref{35}) represents the correction only for the upper half of the drop. It is seen that both (\ref{30}) and (\ref{35}) diverge as $\rho\to R$. So the perturbative solution in the present form fails at the vicinity of the circle $\rho=R$. For $\vartheta > 90^\circ$ one has to find the solution for the convex part (lower half part) as well, represented by the $f_{1-}$. This also can be done in similar lines of the concave part, however the boundary conditions are different. It is easy to see that in this case, just like we had for the upper part, $f_{1-}(\rho)$ diverges for $ \rho\to R $.

One way out of the divergent behavior of $f_{1\pm}(\rho)$ near $\rho=R$ is to change the role of function $z$ and the variable $\rho$, working with $\rho=h(z)$, for $\rho>0$. One can then find valid perturbations near $\rho=R$ (the maximum of $\rho$), however, this time the solution diverges at the top of drop ($\rho\to 0$). This shows that the appearance of the divergent behavior in the perturbative expressions for the shape of drop is not an intrinsic one, but a coordinate artifact. By these all, there are three functions $f_{1\pm}(\rho)$ and $h_1(z)$, which should be joined smoothly to give the correction of the gravity to the shape of the drop in all regions.

A better way is to use the polar coordinate suggested by the sphere solution of the weightless drop.
Choosing the bottom of the circle corresponding to the weightless drop as the origin, and measuring the angle from the $z$-axis, one has  (Fig.~3)
\begin{equation} \label{38}
r (\theta) \cos \theta-d = z = f (\rho) = f_0 (\rho) + \lambda f_1 (\rho)
\end{equation}
\noindent where  $d\equiv R-z_0=R(1+\cos\vartheta) $, by (\ref{13}). A perturbative expansion for $r(\theta)$ is then
\begin{equation}\label{39}
r (\theta) = r_0 (\theta) + \lambda r_1 (\theta)
\end{equation}
where
\begin{equation}\label{40}
 r_0 (\theta) = 2R \cos \theta.
\end{equation}

\begin{figure}[t]
\begin{center}
\includegraphics[width=0.5\columnwidth]{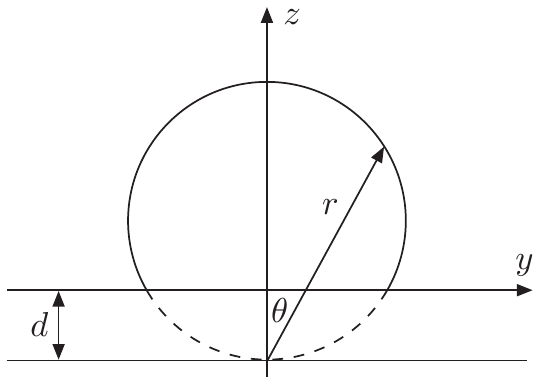}
\caption{The geometry with the polar coordinate. }
\label{fig3}
\end{center}
\end{figure}

\noindent Putting
\begin{equation}\label{41}
\rho = r (\theta) \sin \theta = R \sin (2 \theta) + \lambda r_1 (\theta) \sin \theta
\end{equation}
\noindent in (\ref{38}), one obtains
\begin{align}\label{42}
\lambda ^ 0: & ~~~  r_0 (\theta) \cos \theta-R + z_0 = f_0 (\rho) \Big | _ {\rho = R \sin (2 \theta)} \\
\label{43}
\lambda ^ 1: & ~~~  r_1 (\theta) \cos \theta = \bigg [r_1 (\theta) \sin \theta \; \frac{\partial f_0}{\partial \rho} + f_1 (\rho) \bigg] _ {\rho = R \sin (2 \theta)}.
\end{align}
\noindent As the concave part always exists, we can use $ f_{0+} $ and $ f_{1+} $
in above to find the unknown part $r_1(\theta)$,
\begin{equation}\label{44}
r_1 (\theta) =  \frac{R ^ 3}{V ^ {2 / 3} \cos \theta} \bigg ( \frac{ 3(z_0+a)+2R}{6R}
-\frac{2}{3} \cos (2 \theta) \ln \cos \theta\bigg)
+ b\, \frac{\cos (2 \theta)}{\cos \theta},
\end{equation}
\noindent in which constant $a$ and $b$ are given by (\ref{34}) and (\ref{37}), respectively. In fact the above expression is nothing but the analytically extended result for the concave part (\ref{35}) to the entire surface of the drop. We mention (\ref{44}) has smooth behavior for the whole interval $ 0 \leq \theta   <\frac{\pi}{2} $, which covers the convex part too.  It is useful to define the angle $\theta_0$, as the polar angle at which the contact of the drop and the surface takes place. This angle is easily obtained by the following condition:
\begin{equation}\label{45}
d\,\tan\theta_0 = \rho_0 \simeq R\sin\vartheta+\delta\rho_0,
\end{equation}
in which $\delta\rho_0$ is given by (\ref{33}). For spherical solution $\theta_0=\vartheta/2$.

It is a matter of interest to obtain the equatorial radius (maximum bulge) $\rho_1$ for the case with $\vartheta>90^\circ$. It is obvious that by spherical solution  $\rho_1=R$, happening at the angle $\theta_1=45^\circ$. In general, the equator is defined by the condition:
\begin{equation}\label{46}
0=\rho'(\theta_1)=2R\cos(2\theta_1)+
\lambda \left[ \frac{\mathrm{d}}{\mathrm{d}\theta}(\sin\theta\,r_1(\theta))\right]_{\theta=\pi/4}.
\end{equation}
Using (\ref{41}), at the first order of $\lambda$, we find
\begin{equation}\label{47}
\rho_1=R+ \lambda \sin\frac{\pi}{4} r_1\!\left( \frac{\pi}{4} \right)
\end{equation}
The equatorial plane intercepts $z$-axis at
\begin{align}\label{48}
z_1&=r(\theta_1)\cos\theta_1 - d \\
\label{49}
&= z_0+ R \cos(2\theta_1) + \lambda\cos\frac{\pi}{4}\, r_1\!\left(\frac{\pi}{4}\right)
\end{align}
which is easy to find by using (\ref{46}). For later use, the distance between
the apex of drop and the equatorial plane, $\tilde{h}$, is given explicitly
\begin{equation}\label{50}
\tilde{h}=h-z_1= R+\frac{\varrho g}{\gamma} R^3
\left( \frac{ 3(z_0+a)+2R}{6R}  + \frac{2}{3} \ln \frac{\sqrt{2}}{2}\right).
\end{equation}

\section{Comparison with data}
In order to asses the accuracy of the perturbative solution presented in this work, here the outputs of the solution are compared with some available data. However, it would be helpful to begin with a demonstration of the results. In Fig.~4 the sphere and the perturbative solution are plotted for two drops which are equal except in their contact angles. As expected, the apexes of the perturbative solution are lower than the sphere one's, while the contact radii are increased in comparison with the sphere solution. Also for the case with $\vartheta>90^\circ$, the equatorial radius is larger than the radius of the sphere solution.

\begin{figure}[t]
\begin{center}
\includegraphics[width=0.9\columnwidth]{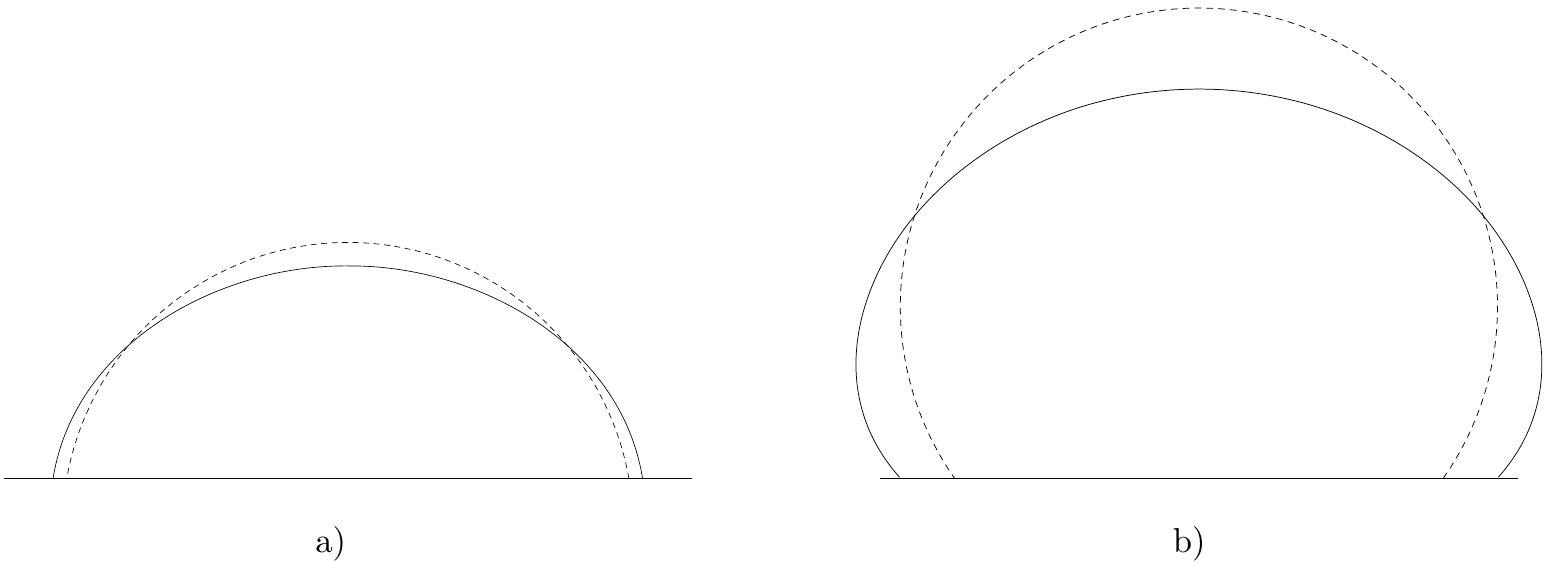}
\caption{Two demonstrations of the sphere solution (dashed line) and the perturbative solution (solid line) for drop a) with acute contact angle ($\vartheta=80^\circ$), and b) with obtuse one ($\vartheta=125^\circ$). For both drops: $V=0.025$~cm$^3$, $\varrho=1$~g/cm$^3$, $\gamma=70$~dyn/cm, $g=980$~cm/s$^2$.}
\label{fig4}
\end{center}
\end{figure}

\begin{table}[t]{\scriptsize
\begin{center}
\begin{tabular}{c c c  c c | c c | c c }
 &  &  cont.  &
\multicolumn{2}{c}{experiment }  &
\multicolumn{2}{c}{sphere sol. (\ref{12})}   &
\multicolumn{2}{c}{perturb. sol. (\ref{39}) }  \\
sample & vol.   &  angle &
$\rho_0$  & $z_1$ &
 $\rho_0$ & $z_1$ &
 $\rho_0$ & $z_1$\\
&  & & & &(err.) &(err.) &(err.) & (err.) \\
\hline & units &&&&&&& \\
& 1 & 72.0$^\circ$ & 0.1748 & 0.1148   & 0.1713 &  0.1245 & 0.1742 &  0.1196 \\
water on&&&&&(2.0\%)&(8.4\%)&(0.3\%)& (4.1\%) \\
carbon  & 2 & 71.3$^\circ$ & 0.2240 & 0.1411   & 0.2171 &  0.1557 & 0.2229 &  0.1460 \\
steel \cite{ryley} & &&&&(3.1\%)&(10\%)&(0.5\%)& (3.4\%) \\
 & 3 & 71.2$^\circ$ & 0.2360 & 0.1565   & 0.2487 &  0.1780 & 0.2575 &  0.1634  \\
&&&&&(5.4\%)&(14\%)&(9.1\%)& (4.4\%) \\
\hline  &cm$^3$&&&&&&& \\
water on &0.1234 & 73.44$^\circ$ &0.4897  & -- & 0.4462 & --  &0.5007 & -- \\
PMMA \cite{kwok} &&&&&(8.9\%)&&(2.3\%)&
\end{tabular}
\caption{{\small The experimental data and theoretical values by spherical solution and the perturbative one for drops of water on two different surfaces. For carbon steel data: 1 unit of vol.=$6.75\times 10^{-3}$~cm$^3$, $\gamma=72$~dyn/cm. For PMMA data: $\gamma=70.6$~dyn/cm. For both surfaces: $\varrho=1$~g/cm$^3$, $g=980.7$~cm/s$^2$. All lengths are in cm.}}
\end{center}
}\end{table}

\begin{table}{\scriptsize
\begin{center}
\begin{tabular}{ccccc| c c c | c c c }
$10^3\times$  &  cont.  &
\multicolumn{3}{c}{experiment }  &
\multicolumn{3}{c}{sphere sol. (\ref{12})}   &
\multicolumn{3}{c}{perturb. sol. (\ref{39}) }  \\
vol.  &  angle &
$\rho_1$ & $\rho_0$  & $\tilde{h}$ &
$\rho_1$ & $\rho_0$  & $\tilde{h}$ &
$\rho_1$ & $\rho_0$  & $\tilde{h}$
\\
cm$^3$  && & & & (err.)& (err.)& (err.)& (err.)&(err.) &(err.) \\ \hline &&&&&&&&&& \\
0.370&131.1$^\circ$&0.0445&0.0337&0.0442&0.0458&0.0345&0.0458&0.0462&0.0357&0.0456 \\
&&&&&(2.8\%)&(2.3\%)&(3.5\%)&(3.9\%)&(5.9\%)& (3.2\%) \\
2.510&129.5$^\circ$&0.0907&0.0722&0.0813&0.0869&0.0671&0.0869&0.0902&0.0748&0.0860 \\
&&&&&(4.2\%)&(7.1\%)&(6.9\%)&(0.6\%)&(3.6\%)& (5.7\%) \\
4.830&132.6$^\circ$&0.1163&0.0884&0.1035&0.1074&0.0791&0.1074&0.1137&0.0957&0.1057 \\
&&&&&(7.6\%)&(11\%)&(3.8\%)&(2.2\%)&(8.3\%)& (2.1\%) \\
10.370&132.4$^\circ$&0.1536&0.1191&0.1299&0.1386&0.1024&0.1386&0.1521&0.1379&0.1349 \\
&&&&&(9.7\%)&(14\%)&(6.7\%)&(1.0\%)&(15.8\%)& (3.9\%) \\
\end{tabular}
\caption{{\small The experimental data and theoretical values by the spherical solution and the perturbative one for drops of mercury on glass slide by \cite{ehrlich}. $\varrho=13.55$~g/cm$^3$, $\gamma=476$~dyn/cm, $g=980.7$~cm/s$^2$.  All lengths are in cm.}}
\end{center}
}\end{table}

The outputs of the perturbative solution are compared with data for drops with both acute and obtuse contact angles. In Table~1 the collection for the case with $\vartheta<90^\circ$ for water drops are presented. The data include three drops on Carbon Steel surface by \cite{ryley}, and one drop on poly(methyl methacrylate) (PMMA) surface by \cite{kwok}. The comparison between the experimental values and the ones by the perturbative solution, perhaps except for the contact radius of third sample on Carbon Steel, shows a satisfactory agreement. The failure of a good agreement with the third drop might be better justified once the peculiar behavior of data with this drop is mentioned. For example, the contact radius of this drop is even less than the one with sphere solution, and the height of apex is larger than the one of sphere solution. Both of these observations are in opposite with expectations from the effect of gravity on a real drop.

In Table~2 the collection for the case with $\vartheta>90^\circ$ for mercury drops are presented. The data include four drops of mercury on glass slide by \cite{ehrlich}. Here the data are available for the equatorial and contact radii, and the equator-apex difference $\tilde{h}$ of the previous section. Again, perhaps except for the first drop, the agreement between experimental data and theoretical results is saisfactory. Just like the third water drop, also for the first drop here the data is in opposite expectations with the effect of gravity (less radii and larger height than the sphere solution). It might be notable that in this collection the Bond parameter, as the perturbative expansion parameter, exceeds one ($\lambda=1.33$) for the last drop as the largest one.

\vskip 0.5cm
\textbf{Acknowledgement}:
This work is supported by the Research Council of the Alzahra University. The comments by A.~Aghamohammadi and specially M.~Khorrami, who pointed to the use of the polar coordinate, are acknowledged.

\end{document}